\documentstyle[html]{article}
\title{Evolution in the Multiverse}
\author{Russell K. Standish\\
High Performance Computing Support Unit\\
University of New South Wales\\Sydney, 2052\\Australia\\
R.Standish@unsw.edu.au\\http://parallel.hpc.unsw.edu.au}
\begin{document}
\maketitle

\begin{abstract}
In the {\em Many Worlds Interpretation} of quantum mechanics, the
range of possible worlds (or histories) provides variation, and the
Anthropic Principle is a selective principle analogous to natural
selection. When looked on in this way, the ``process'' by which the
laws and constants of physics is determined not too different from the
process that gave rise to our current biodiversity, i.e. Darwinian
evolution. This has implications for the fields of SETI and Artificial
Life, which are based on a philosophy of the inevitability of life.
\end{abstract}

\section{Introduction}

The {\em Many Worlds Interpretation} (MWI)\cite{deWitt-Graham73} of
Quantum Mechanics has become increasingly favoured in recent years
over its rivals, with a recent straw poll of eminent
physicists\cite[pp170--1]{Tipler94} showing more than 50\% support it.
David Deutsch\cite{Deutsch97} provides a convincing argument in favour
of MWI, and the {\em multiverse} in the title is due to him.
Tegmark\cite{Tegmark98} has somewhat waggishly suggested that a {\em
  Principle of Plenitude} (alternatively {\em All Universes
  Hypothesis} --- AUH), coupled with the \htmladdnormallinkfoot{{\em
    Anthropic
    Principle}}{http://www.anthropic-principle.com}\footnote{The {\em
    Anthropic Principle} is a statement that the universe we observe
  must be consistent with the existence of us as observers. In the all
  universes hypothesis, the anthropic principle acts to select those
  universes that are ``interesting'', i.e. capable of supporting self
  aware consciousness. In this all universes picture, the distinction
  between the weak and strong forms of the anthropic principle is
  meaningless, so we will simply refer to the Anthropic Principle
  throughout ths paper.}\cite{Barrow-Tipler86} (AP) could be the
ultimate {\em theory of everything} (TOE). Tegmark's Plenitude
consists of all mathematically consistent logical systems, the
principle of plenitude according each of these systems physical
existence, however by the anthropic principle, we should only expect
to find ourselves in a system capable of supporting {\em self-aware
  substructures}, i.e. conciousness. Alternative Plenitudes have
been suggested, for example Schmidhuber's\cite{Schmidhuber97} all
possible programs for a universal turing machine. I have argued
elsewhere\cite{Standish00a}, that the quantum mechanical subset of the
Plenitude, namely the Multiverse, is the most likely system to be
observed by conscious beings.

In this paper, we accept the MWI or Multiverse as a working
hypothesis, and consider what the implications are for evolutionary
systems. An evolutionary system consists of a means of producing
variation, and a means of selecting amongst those variations (natural
selection). Now variations are produced by chance and in the
Multiverse picture, this corresponds to a branching of histories,
whereby a particular entity's offspring will have different forms in
different histories.  The measure of each variant is related to the
proportions in which the variants are formed, and the measure of each
variant evolves in time through a strictly deterministic application
of Schr\"odinger's equation.

What, then, determines which organisms we see today, given that a
priori, any possible history, and hence any mix of organisms may
correspond to our own? Is natural selection completely meaningless?

The first principle we need to apply is the anthropic principle, i.e.
only those histories leading to complex, self-aware substructures will
be selected. We also need to apply the {\em self sampling
  assumption}\cite{Bostrom00a,Bostrom00b} (SSA). The SSA is that each
observer should regard itself as a random sample drawn from the set of
all observer. It is the implicit assumption used in Carter and
Leslie's Doomsday argument\cite{Leslie96}, and much other anthropic
reasoning. Stated another way, as observers, we should expect to see a
world that is nearly maximal in measure, subject to it being
consistent with our existence.  In this picture, natural selection is
a process that differentiates the measure attributed to each variant
organism.

\section{Complexity Growth in Evolution}

As I argued elsewhere\cite{Standish00a}, lawful universes with simple
initial states by far dominate the set consistent with the AP. So the
AP fixes the end point of our evolutionary history (existence of
complex, self-aware organisms), and the SSA fixes the beginning
(evolutionary history is most likely started with the simplest
organisms). We should therefore expect to see an increase in
complexity through time.

What about living systems not governed by the anthropic principle?
Examples include extra terrestrial life (within our own universe, if
it exists) and artificial life systems. Nonhuman terrestrial life is
governed by the AP, since one expects that the evolutionary process
that produced us will also produce the numerous other organisms found
on Earth. A system of life that has evolved completely independently
of Earth has no requirement to produce intelligent beings, and unless
complexity growth is inevitable given the laws of physics and
chemistry, no requirement to produce complex life forms. Proponents of
SETI (the {\em Search for Extra-Terrestrial Intelligence}) believe in
an inevitability of the evolution of intelligent life, given the laws
of physics. The anthropic principle does indeed ensure that the laws
of physics are compatible with the evolution of intelligence, but does
not mandate that this should be likely (excepting, obviously in our
own case). Hanson\cite{Hanson00} has studied a model of evolution
based on easy and hard steps to make predictions about what the
distribution of such steps should be within the fossil record. He
finds that the fossil record is consistent with there being 4--5 hard
steps in getting to intelligent life on Earth. By hard steps, he means
steps who's expected duration greatly exeeds the present age of the
universe. The hard steps include
\begin{itemize}
\item origin of first replicator
\item origin of sex
\item origin of eukaryotic cells
\item origin of multicellularity
\item possibly the origin of self-aware conscious entities
\end{itemize}
This would imply that intelligent life is fairly unique
within our own universe, to the chagrin of the SETI proponents,
but simple prokaryotic life may well be ubiquitous. Of
course, it is also true that a single example of extra terrestrial
intelligence would be an important counterexample to these arguments
based on the AP and SSA, so SETI is by itself not a fruitless
exercise. 


Likewise, for artificial life, it would seem plausible that a serious
of easy and hard steps are required to climb the complexity ladder.
Already, the first such hard transition (the creation of replicators
from the primeval soup) has been
observed\cite{Pargellis96a,Pargellis96b}, but equivalents of other
transitions (eg transition to sexual reproduction, prokaryote to
eukaryote or multicellularity) have not been observed to date. Ray is
leading a major experiment designed to probe the transition to
multicellularity\cite{Ray??,Ray-Hart98} --- success in this experiment
will provide remarkable constraints on just how finely tuned the
physics and chemistry needs to be in order for the system to pass
through a hard transistion.

Adami\cite{Adami98a,Adami99a} and co-workers examined the {\em Avida}
alife system for evidence of complexity growth during evolution. They
did find this, although this is largely seen as the artificial
organisms learning how to solve arithmetic problems that have been
imposed artificially on the system. An analogous study by
myself\cite{Standish99a} of Tierra showed no such increase in
complexity over time --- if anything the trend was to greater
simplicity. This work is still in progress.

\section{Evolutionary Physics?}

Returning back to the picture of the ``All Universes Hypothesis'', we
can see that our current universe is made up from contingency and
necessity. The necessity comes from the requirements of the anthropic
principle, however when a particular aspect of the universe is not
constrained by the AP, its value must be decided by chance (according
to the SSA) the first time it is ``measured'' by self-aware beings
(this measurement may well be indirect --- properties of the
microscopic or cosmic worlds will need to be consistent with our
everyday observations at the macroscopic level, so may well be
determined prior to the first direct measurements). Evolution is also
described as a mixture of contingency and necessity. When understood
in terms of the AP supplying the necessary, and the SSA supplying the
rationale for resolving chance, the connection between the selection
of phyical laws and the selection of organisms in evolution is made
clear. It is as though the laws of physics and chemistry have
themselves evolved. Perhaps applying evolutionary principles to the
underlying physico-chemical laws of an alife system will result in an
alife system that can pass through these hard transitions.

\bibliographystyle{plain}
\bibliography{rus}

\end{document}